\documentclass[conference,a4paper,final]{IEEEtran}

\IEEEoverridecommandlockouts
\usepackage{cite}
\usepackage{float}
\usepackage[printonlyused,nolist]{acronym}
\usepackage{subcaption}
\usepackage{amsmath,amssymb,amsfonts}
\usepackage{amsthm}
\usepackage{makecell}
\usepackage{tabularx}
\usepackage{amscd}
\usepackage{graphicx}
\usepackage{color}
\usepackage{enumitem}
\usepackage{amsfonts}
\usepackage{textcomp}
\usepackage{booktabs}
\usepackage[export]{adjustbox}
\usepackage{dblfloatfix}
\usepackage[utf8]{inputenc}

\usepackage{verbatim} 
\usepackage[hidelinks, bookmarks=false]{hyperref}
\hypersetup{draft}
\usepackage[T1]{fontenc}
\usepackage{algorithm}
\usepackage{algorithmic}
\usepackage{theoremref}
\hypersetup{
pdftitle={Power and Beam Optimization for Uplink Millimeter-Wave Hotspot Communication Systems},
pdfsubject={5G mmWave Wireless Networks},
pdfauthor={Rafail Ismayilov, Bernd Holfeld, Renato L. G. Cavalcante and Megumi Kaneko},
pdfkeywords={Radio resource management, Interference management, Millimeter-wave networks, Power optimization, Beamforming, 5G, Uplink}
}

\def\BibTeX{{\rm B\kern-.05em{\sc i\kern-.025em b}\kern-.08em
    T\kern-.1667em\lower.7ex\hbox{E}\kern-.125emX}}

\newtheoremstyle{dotless}{}{}{\itshape}{}{\bfseries}{}{ }{}
\theoremstyle{dotless}
\newtheorem{definition}{Definition}

\newtheorem{problem}{Problem}

\newcommand{\overbar}[1]{\mkern 2.6mu\overline{\mkern-1.5mu#1\mkern-1.5mu}\mkern 1.5mu}

\DeclareMathOperator*{\argmax}{arg\,max}
\DeclareMathOperator*{\argmin}{arg\,min}

\def\tmin{{\text{min}}}
\def\tmax{{\text{max}}}

\def\setN{{\mathcal{N}}}

\def\setM{{\mathcal{M}}}

\def\setQ{{\mathcal{Q}}}

\def\pwrv{{\boldsymbol{p}}}
\def\bw{{\boldsymbol{\theta}}}
\def\bd{{\boldsymbol{\beta}}}
\def\pR{\mathbb{R}_{++}}
\def\nnR{\mathbb{R}_{+}}

\makeatletter
\newcommand\fs@betterruled{%
  \def\@fs@cfont{\bfseries}\let\@fs@capt\floatc@ruled
  \def\@fs@pre{\vspace*{5pt}\hrule height.8pt depth0pt \kern2pt}%
  \def\@fs@post{\kern2pt\hrule\relax}%
  \def\@fs@mid{\kern2pt\hrule\kern2pt}%
  \let\@fs@iftopcapt\iftrue}
\floatstyle{betterruled}
\restylefloat{algorithm}
\makeatother

\begin{document}
\renewcommand{\baselinestretch}{0.95}
\title{Power and Beam Optimization for Uplink Millimeter-Wave Hotspot Communication Systems
\vspace{-0.4em}}
\author{
	\IEEEauthorblockN{Rafail Ismayilov\IEEEauthorrefmark{1}, Bernd Holfeld\IEEEauthorrefmark{1}, Renato L. G. Cavalcante\IEEEauthorrefmark{1}\IEEEauthorrefmark{2} and Megumi Kaneko\IEEEauthorrefmark{3}}
	\IEEEauthorblockA{\IEEEauthorrefmark{1}Fraunhofer Heinrich Hertz Institute, Berlin, Germany}
	\IEEEauthorblockA{\IEEEauthorrefmark{2}Technical University of Berlin, Berlin, Germany}
	\IEEEauthorblockA{\IEEEauthorrefmark{3}National Institute of Informatics, Tokyo, Japan}
	\vspace{-0.4em}
	\\ E-mail: \{rafail.ismayilov, bernd.holfeld, renato.cavalcante\}@hhi.fraunhofer.de, megkaneko@nii.ac.jp}
\maketitle


\begin{abstract}
We propose an effective interference management and beamforming mechanism for uplink communication systems that yields fair allocation of rates.
In particular, we consider a hotspot area of a \ac{mmWave} access network consisting of multiple \ac{UE} in the uplink and multiple \acp{AP} with directional antennas and adjustable beam widths and directions (beam configurations).
This network suffers tremendously from multi-beam multi-user interference, and, to improve the uplink transmission performance, we propose a centralized scheme that optimizes the power, the beam width, the beam direction of the \acp{AP}, and the \ac{UE} - \ac{AP} assignments.
This problem involves both continuous and discrete variables, and it has the following structure. 
If we fix all discrete variables, except for those related to the \ac{UE}-\ac{AP} assignment, the resulting optimization problem can be solved optimally.
This property enables us to propose a heuristic based on \ac{SA} to address the intractable joint optimization problem with all discrete variables.
In more detail, for a fixed configuration of beams, we formulate a weighted rate allocation problem where each user gets the same portion of its maximum achievable rate that it would have under non-interfered conditions.
We solve this problem with an iterative fixed point algorithm that optimizes the power of \acp{UE} and the \ac{UE} - \ac{AP} assignment in the uplink.
This fixed point algorithm is combined with \ac{SA} to improve the beam configurations.
Theoretical and numerical results show that the proposed method improves both the \ac{UE} rates in the lower percentiles and the overall fairness in the network.
\end{abstract} 
\vspace{0.2cm}

\acresetall 
\section{Introduction}\label{s1}
\IEEEPARstart{T}{o} support huge data rates in next-generation communication systems, \ac{mmWave} technologies using wideband signals are widely considered as an attractive technology \cite{Niu2015}.
From a research perspective, one of the challenges to overcome is the high \ac{PL} of the \ac{mmWave} band compared to that of traditional bands.
The channel \ac{PL} in the \ac{mmWave} bands is generally higher than that of traditional frequencies \cite{Rappaport2013}.
In particular, the inherent propagation characteristics make the use of \ac{mmWave} transmission sensitive to blockage.
Thus, \ac{MIMO} and \ac{BF} techniques are adopted to compensate the severe \ac{PL} conditions \cite{Kutty2016, Han2017}.
Directional transmission is also known to be beneficial for reducing the interference in networks and for improving the spatial reuse of radio resources and the transmission range.
However, with the densification of networks, directional transmission with narrow beams creates additional difficulties.

\begin{figure}[t]
	\centering
	\includegraphics[width=78mm]{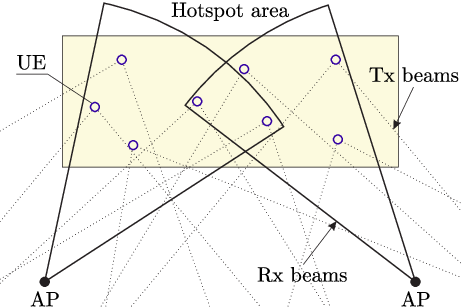}
	\caption{A \ac{mmWave} network with transmit and receive beamforming for uplink connectivity in areas of high user density, where interference management is a key challenge.}
	\label{fig1}
	\vspace{-10pt}
\end{figure}

In contrast to traditional \ac{RRM} with physical (PHY) and \ac{MAC} cross-layer approaches, where resources are usually managed in a time-frequency domain, \ac{mmWave} communication systems also need to select appropriate transmit and receive beam directions and widths (beam configurations) of the network entities.
As illustrated in Fig.~\ref{fig1}, the large numbers of \ac{UE} connected to different \acp{AP} would have to share the frequency resources in the uplink, thus interference managament schemes coping with mutual interference are required.
However, the design of interference management schemes that jointly optimizes the power, \ac{UE}-\ac{AP} assignments, and the beam configuration is difficult. 

In this study, we consider a wireless \ac{mmWave} access network where multiple low-mobility users in a hotspot area communicate with a set of \acp{AP} in the uplink using optimized beam steering. 
We start by analyzing the possible interference cases in the considered multi-beam multi-user scenario.
Next, for a given power budget of \acp{UE} and discrete beam configurations of \acp{AP} in the network, we pose a weighted max-min problem involving the joint optimization of power and \ac{UE} - \ac{AP} assignments in the uplink. 
We show that this problem can be easily solved with a simple fixed point algorithm that is further combined with a heuristic based on simulated annealing \cite{haggstrom2002finite} to search for an optimal beam configuration. 

Our work builds upon a previous study on throughput and fairness trade-offs depending on beam width selection in multi-beam multi-user \ac{mmWave} communication systems \cite{Rafail2018}.
Interference management via transmit beam width and direction for improving the system performance is one of the center topics in \ac{mmWave} communications. 
E.g., the authors in \cite{Onireti2018} consider uplink \ac{mmWave} cellular networks and minimize the interference by adapting only the transmit power of the \acp{UE}.
\cite{Luo2017} proposes a performance optimization approach for uplink \ac{mmWave} communication systems based on a spatial modulation scheme.
This scheme assumes an exact orthogonality between different beams, and such assumption is not valid for \ac{mmWave} hotspot networks.
Moreover, the impact of the transmit and receive beam widths to the system performance was not studied.
Uplink inter-user interference in \ac{mmWave} systems was considered in \cite{Li2016}.
The proposed scheme takes into consideration a single-cell scenario and assumes that the \ac{CSI} is known perfectly at the \ac{AP}. 

\section{Preliminaries} \label{s3}
In this study, we use the following standard definitions:
scalars and variables are denoted by lowercase letters (e.g. $x$ and $y$).
We use boldface letters to emphasize vectors (e.g. $\boldsymbol{x}$ and $\boldsymbol{y}$).
The $i$th element of a vector $\boldsymbol{x}$ is denoted by $x_i$.
A vector inequality $\boldsymbol{x} \le \boldsymbol{y}$ should be understood as an element-wise inequality.
Sets are defined with calligraphic fonts (e.g. $\mathcal{X}$ and $\mathcal{Y}$).
Probability distributions are denoted with calligraphic letters.
By $\left \| \cdot \right \|_{\infty }$, we denote the standard $l-\infty$ norm.
Sets of non-negative and positive reals are denoted by $\nnR$ and $\pR$, respectively.

\subsection{Uplink Network Model} \label{s3a}
We consider a wireless network comprised of a set $\setN = \{1,...,N\}$ of transmitters (Tx), called \acf{UE}, and a set $\setM = \{1,...,M\}$ of receivers (Rx), called \acfp{AP}.
We assume fixed transmit beam widths $\theta^{\text{Tx}}_n = \theta^{\text{Tx}}, \forall n \in \setN$.
The transmit beam directions of the \acp{UE} are uniformly distributed with $\beta^{\text{Tx}}_n \sim \mathcal{U}(\beta^{\text{Tx}}_{\text{min}},\beta^{\text{Tx}}_{\text{max}}), \forall n \in \setN$.
Furthermore, we assume a transmit power constraint for each \ac{UE} given by $\overbar{P}$.
The transmit power vector $\pwrv = \left ( p_1,...,p_N \right ) \in \nnR^N$ takes values from a continuous power domain; i.e., $p_n \in \nnR$, $p_n \le \overbar{P}$, $\forall n \in \mathcal{N}$.
In contrast to the transmitter side, we assume that each receive beam width and direction can be adjusted by the \ac{AP}.
Let $\bw = \left ( \theta^{\text{Rx}}_1,...,\theta^{\text{Rx}}_M \right ) \in \mathbb{D}^M_{\theta} \subseteq \pR^M$ be the receive beamwidth vector, where $\theta^{\text{Rx}}_m$ takes values from a discrete set $\mathbb{D}_{\theta}$ and $\theta_{\tmin}^{\text{Rx}} \le \theta^{\text{Rx}}_m \le \theta_{\tmax}^{\text{Rx}}$, $\forall m \in \setM$.
Similarly, each receive beam can be steered by the \ac{AP} in a specific angular direction and the vector of receive beam directions is denoted by $\bd = \left ( \beta_1^{\text{Rx}},...,\beta_M^{\text{Rx}} \right ) \in \mathbb{D}^M_{\beta} \subseteq \pR^M$, where $\beta_m^{\text{Rx}}$ takes values from a discrete set $\mathbb{D}_{\beta}$ and $\beta^{\text{Rx}}_{\text{min}} < \beta_m^{\text{Rx}} < \beta^{\text{Rx}}_{\text{max}}$, $\forall m \in \setM$.

In this work, we assume that multiple \acp{UE} may be simultaneously connected to an \ac{AP}, constituting a many-to-one association scenario.
Hence, the \ac{AP} is capable of processing several incoming uplink signals at the same time.

The \ac{CSI}, which is needed to perform the beam and interference management, is assumed to be composed of the large-scale channel fading gains, based on the \ac{mmWave} \ac{PL} model (see Section \ref{s7a}), for all users. 
That is, the instantaneous small-scale fading coefficients are assumed to be unknown, otherwise this would generate an excessively high amount of \ac{CSI} feedback overhead, hardly implementable in \ac{mmWave} systems. 

Fig.~\ref{fig1} illustrates an example of a hotspot scenario where \acp{UE} are randomly and uniformly distributed with high density.

\begin{figure}[t]
	\centering
	\includegraphics[width=90mm]{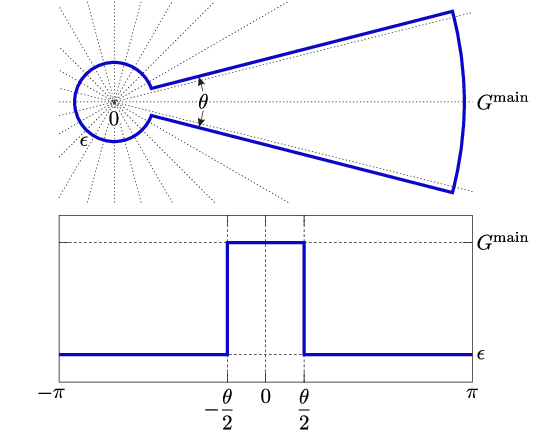}
	\caption{Model of a symmetric sector antenna pattern with beam width $\theta$ and beam gains $G^{\text{main}}$ in the mainlobe and $\epsilon$ in the sidelobe. \vspace{-10pt}}
	\label{fig2}
\end{figure}

\begin{figure*}[ht]
	\centering
	\begin{subfigure}{0.24\textwidth}
		\includegraphics[width=43mm,center]{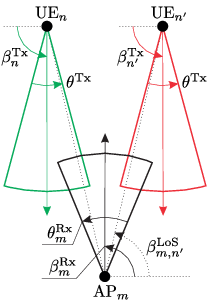}
		\caption{$\begin{array}{ll}
			G^{\text{Tx}}_{n'} = G^{\text{Tx,main}}_{n'}
			\vspace{4pt}
			\\
			G^{\text{Rx}}_m = G^{\text{Rx,main}}_m
			\end{array}$}
		\label{fig3a}
	\end{subfigure}
	\begin{subfigure}{0.24\textwidth}
		\includegraphics[width=43mm,center]{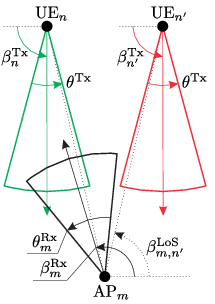}
		\caption{$\begin{array}{ll}
			G^{\text{Tx}}_{n'} = G^{\text{Tx,main}}_{n'}
			\vspace{4pt}
			\\
			G^{\text{Rx}}_m = \epsilon
			\end{array}$}
		\label{fig3b}
	\end{subfigure}
	\begin{subfigure}{0.24\textwidth}
		\includegraphics[width=43mm,center]{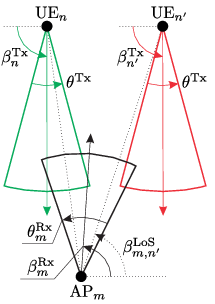}
		\caption{$\begin{array}{ll}
			G^{\text{Tx}}_{n'} = \epsilon
			\vspace{4pt}
			\\
			G^{\text{Rx}}_m = G^{\text{Rx,main}}_m
			\end{array}$}
		\label{fig3c}
	\end{subfigure}
	\begin{subfigure}{0.24\textwidth}
		\includegraphics[width=43mm,center]{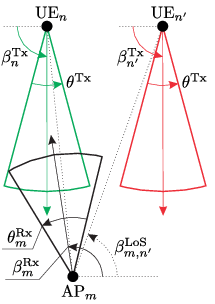}
		\caption{$\begin{array}{ll}
			G^{\text{Tx}}_{n'} = \epsilon
			\vspace{4pt}
			\\
			G^{\text{Rx}}_m = \epsilon
			\end{array}$}
		\label{fig3d}
	\end{subfigure}
	\caption{Considered interference cases: UE$_{n'}$ causes interference in the transmit mainlobe of UE$_n$ which is connected to AP$_m$}
	\label{fig3}
\end{figure*}
\begin{figure*}[!hb]
	\hrule
	\vspace{8pt}
	\begin{equation} \label{eq8}\tag{8}
	\resizebox{0.95\hsize}{!}{$
        G^{\text{Tx}}_{n'}(\theta_{n'}^{\text{Tx}},\beta_{n'}^{\text{Tx}}) \, G^{\text{Rx}}_{m}(\theta_{m}^{\text{Rx}},\beta_{m}^{\text{Rx}}) = \left\{
        \begin{array}{lllll}
		G^{\text{Tx,main}}_{n'} \, G^{\text{Rx,main}}_m &, \text{if} & 0 < \begin{vmatrix} \beta^{\text{LoS}}_{m,n'} - \beta^{\text{Tx}}_{n'} \end{vmatrix} <  \dfrac{\theta^{\text{Tx}}}{2} &\text{and} & 0 < \begin{vmatrix} \beta^{\text{LoS}}_{m,n'} - \beta^{\text{Rx}}_m \end{vmatrix} <  \dfrac{\theta^{\text{Rx}}_{m}}{2}  \vspace{6pt}
		\\
		\epsilon \, G^{\text{Tx,main}}_{n'} &, \text{if} & 0 < \begin{vmatrix} \beta^{\text{LoS}}_{m,n'} - \beta^{\text{Tx}}_{n'} \end{vmatrix} <  \dfrac{\theta^{\text{Tx}}}{2} & \text{and} & \dfrac{\theta^{\text{Rx}}_{m}}{2} < \begin{vmatrix} \beta^{\text{LoS}}_{m,n'} - \beta^{\text{Rx}}_m \end{vmatrix}
		\vspace{6pt}
		\\
		\epsilon \, G^{\text{Rx,main}}_m &, \text{if} & \dfrac{\theta^{\text{Tx}}}{2} < \begin{vmatrix} \beta^{\text{LoS}}_{m,n'} - \beta^{\text{Tx}}_{n'} \end{vmatrix} & \text{and} & 0 < \begin{vmatrix} \beta^{\text{LoS}}_{m,n'} - \beta^{\text{Rx}}_m \end{vmatrix} <  \dfrac{\theta^{\text{Rx}}_{m}}{2}
		\vspace{6pt}
		\\
		\epsilon^2 &, \text{otherwise} & & &
		\end{array}\right.
		$}
	\end{equation}
\end{figure*}

\subsection{Directive Antenna Patterns} \label{s3b}
The beam width is one of the key variables that we will adjust in the proposed scheme in order to improve the system performance.
We refer to an antenna model presented in \cite{Shokri2015}.
It uses the simplified and approximated beam gain pattern provided in Fig.~\ref{fig2} for both transmitters and receivers.
An antenna with a gain pattern defined by beam width $\theta \in (0,2\pi)$, gain in the mainlobe $G^{\text{main}}$, and gain in the sidelobe $\epsilon$ with $0 < \epsilon < 1 < G^{\text{main}}$ can be expressed by
\begin{equation}\label{eq1}
\resizebox{0.91\hsize}{!}{$
	G(\gamma) = \left\{\begin{array}{ll}
	G^{\text{main}} = \dfrac{2\pi - (2\pi - \theta)\epsilon}{\theta} &, \text{if} \hspace{5pt} |\gamma| \le \dfrac{\theta}{2} \\
	\epsilon &, \text{otherwise.}
	\end{array}\right.
	$}
\end{equation}
Obviously, the beam gains in the mainlobe are increasing with smaller beamwidth.
With $\theta = 2 \pi$ we have an omnidirectional mode with unit gain.


\subsection{Interference Model} \label{s3d} %
We adopt the interference model studied in \cite{Xue2017, Zhang2018}.
An \ac{UE} $n \in \setN$ is connected to a single \ac{AP} $m \in \setM$, and the radiated power from other \acp{UE} $n' \neq n$ is treated as the interference power at the \ac{AP} $m$.
Hence, the overall interference at the receiver is expressed by
\vspace{-5pt}
\begin{equation}\label{eq4}
	I_{m,n'} =
	\sum_{\substack{n' \in \setN \setminus\left \{ n \right \}}}^{ } p_{n'}h_{m,n'}(\theta_{n'}^{\text{Tx}},\beta_{n'}^{\text{Tx}},\theta_{m}^{\text{Rx}},\beta_{m}^{\text{Rx}}),
\end{equation}
where $p_{n'}$ is the transmit power of the interfering \ac{UE} $n' \neq n$ and $h_{m,n'}(\theta_{n'}^{\text{Tx}},\beta_{n'}^{\text{Tx}},\theta_{m}^{\text{Rx}},\beta_{m}^{\text{Rx}})$ is the power gain of the interference channel between \ac{UE} $n'$ and \ac{AP} $m$.
The latter depends on Tx beamwidth $\theta_{n'}^{\text{Tx}}$, Tx beam direction $\beta_{n'}^{\text{Tx}}$, Rx beamwidth $\theta_{m}^{\text{Rx}}$, Rx beam direction, $\beta_{m}^{\text{Rx}}$ and the distance from \ac{AP} $m$ to \ac{UE} $n'$.
The interference power gain is expressed as follows:
\begin{equation} \label{eq5}
\resizebox{0.91\hsize}{!}{$
	\hspace{-5pt}
    h_{m,n'}(\theta_{n'}^{\text{Tx}},\beta_{n'}^{\text{Tx}},\theta_{m}^{\text{Rx}},\beta_{m}^{\text{Rx}}) = G^{\text{Tx}}_{n'}(\theta_{n'}^{\text{Tx}},\beta_{n'}^{\text{Tx}})
    \hspace{2pt}
    G^{\text{Rx}}_{m}(\theta_{m}^{\text{Rx}},\beta_{m}^{\text{Rx}})
    \hspace{2pt} \text{PL}_{m,n'},
    \hspace{-5pt}
$}
\end{equation}
where $G^{\text{Tx}}_{n'}(\theta_{n'}^{\text{Tx}},\beta_{n'}^{\text{Tx}})$ and $G^{\text{Rx}}_{m}(\theta_{m}^{\text{Rx}},\beta_{m}^{\text{Rx}})$ are transmit and receive beam gains of \ac{UE} $n'$ and \ac{AP} $m$, respectively.
The scalar $\text{PL}_{m,n'}$ denotes the path loss between \ac{UE} $n'$ and \ac{AP} $m$.
As mentioned in Section~\ref{s3a}, we assume that the transmit beam width of all \acp{UE} is fixed $(\theta^{\text{Tx}}_n = \theta^{\text{Tx}}, \forall n \in \setN)$, and \ac{UE} $n$ is always in \ac{LoS} with \ac{AP} $m$, if this is its serving access point.

We distinguish four interference scenarios, as shown in Fig.~\ref{fig3} (a)-(d).
The respective transmit and receive beam gains are calculated as follows:
\begin{equation} \label{eq6}
\resizebox{0.91\hsize}{!}{$
    G^{\text{Tx}}_{n'}(\theta_{n'}^{\text{Tx}},\beta_{n'}^{\text{Tx}}) = \left\{\begin{array}{ll}
	G^{\text{Tx,main}}_{n'} &, \text{if} \hspace{5pt} 0 < \begin{vmatrix} \beta^{\text{LoS}}_{m,n'} - \beta^{\text{Tx}}_{n'} \end{vmatrix} < \dfrac{\theta^{\text{Tx}}}{2} \\
	\epsilon  &, \text{otherwise}
	\end{array}\right.
	$}
\end{equation}
\begin{equation} \label{eq7}
\resizebox{0.91\hsize}{!}{$
    G^{\text{Rx}}_{m}(\theta_{m}^{\text{Rx}},\beta_{m}^{\text{Rx}}) = \left\{\begin{array}{ll}
	G^{\text{Rx,main}}_{m} &, \text{if} \hspace{5pt} 0 < \begin{vmatrix} \beta^{\text{LoS}}_{m,n'} - \beta^{\text{Rx}}_m \end{vmatrix} < \dfrac{\theta^{\text{Rx}}_{m}}{2} \\
	\epsilon  &, \text{otherwise.}
	\end{array}\right.
	$}
\end{equation}
Above, $G^{\text{Tx,main}}_{n'}$ and $G^{\text{Rx,main}}_{m}$ denote the mainlobe gains of \ac{UE} $n'$ and \ac{AP} $m$ according to \eqref{eq1}.
For all four interference cases, expression \eqref{eq8} gives the combined transmit and receive beam gains that can be obtained in the network.
\addtocounter{equation}{1}
\section{Problem Statement} \label{s5}
The objective of this study is to maximize the system utility in the network, which we define as a weighted rate allocation problem.
The problem involves the optimization of the \ac{UE}-\ac{AP} assignments, the receive beam widths $(\bw)$, the receive beam directions $(\bd)$, and the transmit power $(\pwrv)$.
In addition, the possible beam configurations are subject to discrete candidate sets $\mathbb{D}_{\theta}$ and $\mathbb{D}_{\beta}$, and each component of the power vector $\pwrv$ cannot exceed the value $\overbar{P} \in \pR$.
\vspace{-5pt}
\subsection{Uplink Data Rates} \label{s5a}
For $\pwrv$, $\bw$ and $\bd$ given, the \ac{SINR} at \ac{AP} $m \in \setM $ is defined as follows:
\begin{equation}\label{eq9}
	\begin{array}{rcl}
		\hspace{-5pt}s_{n}:\nnR^N \times \pR^M \times \pR^M \times \setM & \rightarrow & \nnR                                                     \\
		(\pwrv, \bw, \bd, m)                                                  & \mapsto     & \dfrac{p_n \hspace{1pt} h_{m,n}}{I_{m,n'} + \sigma^2_{\text{noise}}},
	\end{array}
\end{equation}
where $p_n$ is the transmit power of \ac{UE} $n \in \setN$ being connected to \ac{AP} $m \in \setM$.
The term $I_{m,n'}$ is the interference power defined in \eqref{eq4}, $\sigma^2_{\text{noise}}$ is the noise power at all \acp{AP} (which we assume to be equal) and $h_{m,n}$ refers to the channel power gain between the serving \ac{AP} and the \ac{UE}, given by
\begin{equation*}
	h_{m,n} = G^{\text{Tx,main}}_n\hspace{2pt} G^{\text{Rx,main}}_m\hspace{2pt} \text{PL}_{m,n}.
\end{equation*}
Above, $G^{\text{Tx,main}}_n$ and $G^{\text{Rx,main}}_m$ are transmit and receive beam gains in the mainlobe of \ac{UE} $n \in \setN$  and \ac{AP} $m \in \setM$, respectively, and $\text{PL}_{m,n}$ is the path loss.

Hereafter, the \emph{achievable rate} in the uplink of \ac{UE} $n \in \setN $ to its best serving \ac{AP} (e.g., \ac{UE}-\ac{AP} assignment) is expressed by
\begin{equation}\label{eq10}
	R_n(\pwrv,\bw,\bd) = \underset{m \in \setM}{\text{max}} W\log_2\begin{pmatrix}1+s_n(\pwrv,\bw,\bd,m)\end{pmatrix},
\end{equation}
where $W$ is the system bandwidth.
For $p_n$ being fixed to the maximum transmit power budget $\overbar{P} > 0$, the maximum achievable rate, called \emph{interference-free rate}, is given by
\begin{equation} \label{eq13}
	\overbar{R}_n = \underset{m \in \setM}{\text{max}} W \log_2 \left ( 1+\dfrac{\overbar{P} \hspace{2pt} h_{m,n}}{\sigma^2_{\text{noise}}} \right ).
\end{equation}
In other words, $\overbar{R}_n$ is the rate corresponding to the case of \ac{UE} $n \in \setN$ transmitting alone in the network with full power to its best serving \ac{AP}.

\subsection{The Weighted Rate Allocation Problem} \label{s5b}
\begin{figure}[t]
	\centering
	\includegraphics[width=79mm]{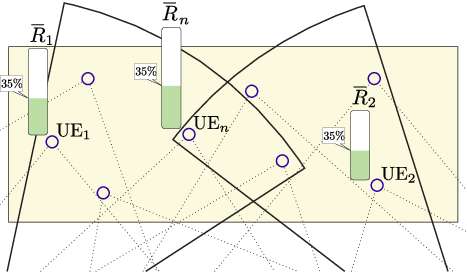}
	\caption{Illustration of the weighted rate allocation scheme. Each user gets a portion $c=35\%$ of its interference-free rate.}
	\label{fig5}
\end{figure}

As illustrated in Fig.~\ref{fig5}, the objective of the optimization problem is to assign the user rates $R_n$, $\forall n \in \setN$, fairly, in the sense that every \ac{UE} $n \in \setN$ achieves the maximum \emph{common fraction} $c \in [0,1]$ of the interference-free rates $\overbar{R}_n$.
Formally, the proposed optimization problem is stated as the following mixed integer problem:
\vspace{-5pt}
\begin{subequations} \label{eq11}
	\begin{alignat}{3}
		  & \underset{\pwrv,\bw,\bd,c}{\text{\normalfont{maximize}}} & \quad & c \tag{\ref{eq11}}                    \\
		  & \text{\normalfont{subject to}}                   &       & c\overbar{R}_n = R_n(\pwrv,\bw,\bd), \, \forall n \in \setN   \label{eq11a}                                                    \\
		  &                                     &       & \left \| \pwrv \right \|_\infty  \le \overbar{P}                                                                          \label{eq11b} \\
		  &                                     &       & \pwrv \in \nnR^N, (\bw,\bd) \in \setQ, c \in \pR ,   \label{eq11c}
	\end{alignat}
\end{subequations}
where $\pwrv$ is the transmit power vector, $\overbar{P}$ is the power budget, and $\setQ = \left \{ (\bw_k,\bd_l)\right \}_{\substack{k \in (1,...,\left | \mathbb{D}_{\theta} \right |^M), \hspace{2pt} l \in (1,...,\left | \mathbb{D}_{\beta} \right |^M)}}$ is a set of all beam configurations of the \acp{AP}.

Owing to the discrete parameters, it is hard to solve problem \eqref{eq11}.
However, if the tuple $(\bw,\bd)$ is fixed to a given beam configuration $(\overbar{\bw},\overbar{\bd}) \in \setQ$, one can optimally solve the weighted rate allocation problem \eqref{eq11}. 
In this case, the objective reduces to the following problem:
\vspace{-5pt}
\begin{subequations} \label{eq11plus}
	\begin{alignat}{3}
		  & \underset{\text{s.t. } (\pwrv,c) \in \mathcal{K}_{(\overbar{\bw},\overbar{\bd})}}{\text{\normalfont{maximize}}} & \quad & c \tag{\ref{eq11plus}} , 	\end{alignat}
\end{subequations}
\noindent where $\mathcal{K}_{(\overbar{\bw},\overbar{\bd})}$ is the set of constraints \eqref{eq11a}-\eqref{eq11c} excluding the constraint on the beam configurations, which are fixed to $(\overbar{\bw},\overbar{\bd})$.
Problem \eqref{eq11plus} can be efficiently solved with an iterative fixed point algorithm that is described in the next section and also in the Appendix.

Problem \eqref{eq11plus} also enables us to define a function that maps an arbitrary beam configuration $(\bw,\bd) \in \setQ$ to a rate fraction $c^{\star}$ as follows:
\begin{equation}\label{utility_map}
	U:\setQ \rightarrow \pR:(\bw,\bd) \mapsto c^{\star},
	\end{equation}	
where $c^{\star}$ is the component of the tuple ($\pwrv^{\star},c^{\star}$) that solves \eqref{eq11plus} for a given beam configuration $(\overbar{\bw},\overbar{\bd})$.
In this study, we propose to maximize $U$ with a \ac{SA} algorithm that adjusts the receive beam width and direction of the \acp{AP}.
Briefly, the proposed \ac{SA} approach selects the parameters $(\bw^\star,\bd^\star)$ from a discrete beam configuration set $Q$ such that
\begin{equation}\label{eq_utility}
(\bw^\star,\bd^\star) \in \underset{(\bw,\bd) \in \setQ}{{\argmax}} \hspace{3pt} U(\bw,\bd),
\end{equation}
and the remaining optimal variables of problem \eqref{eq11plus} are obtained with the fixed point algorithm described next.

\section{Solution Framework} \label{s6}
\subsection{Optimal Utility Power Allocation, and \ac{UE}-\ac{AP} Assignment for a given Beam Configuration} \label{s6a}
To reformulate problem \eqref{eq11} in the canonical form \eqref{eq18} in the Appendix for a given $(\overbar{\bw},\overbar{\bd})$, we first apply the following transformation for every $n \in \setN$ (assuming $c>0$):
\begin{equation*}
\resizebox{1.0\hsize}{!}{$
	\begin{array}{rl}
		c\overbar{R}_n = R_n(\pwrv,\overbar{\bw},\overbar{\bd})  \Leftrightarrow & c\overbar{R}_n = \underset{m \in \setM}{\text{max}} W\log_2\begin{pmatrix}1+s_n(\pwrv,\overbar{\bw},\overbar{\bd},m)\end{pmatrix} \vspace{5pt}                     \\
		\Leftrightarrow                   & \dfrac{1}{c} = \underset{m \in \setM}{\text{min}}\dfrac{\overbar{R}_n}{W\log_2\begin{pmatrix}1+s_n(\pwrv,\overbar{\bw},\overbar{\bd},m)\end{pmatrix}} \vspace{5pt} \\
		\Leftrightarrow                   & \dfrac{1}{c}p_n = \underset{m \in \setM}{\text{min}}\dfrac{\overbar{R}_n p_n}{W\log_2\begin{pmatrix}1+s_n(\pwrv,\overbar{\bw},\overbar{\bd},m)\end{pmatrix}},
	\end{array}
	$}
\end{equation*}
Or, more compactly,
\vspace{-1pt}
\begin{equation} \label{eq14}
	\pwrv \in \text{Fix} \left (cT^{(\overbar{\bw},\overbar{\bd})} \right),
	\vspace{-1pt}
\end{equation}
\noindent where 
\vspace{-1pt}
\begin{equation} \label{eq15}
	T^{(\overbar{\bw},\overbar{\bd})}:\nnR^N \rightarrow \pR^N: \pwrv \mapsto \left [ t_1^{(\overbar{\bw},\overbar{\bd})}(\pwrv), ...,t_N^{(\overbar{\bw},\overbar{\bd})}(\pwrv) \right ],
	\vspace{-1pt}
\end{equation}
and 
\begin{equation*}
\resizebox{1.0\hsize}{!}{$
	\begin{array}{rcl}
		t_n^{(\overbar{\bw},\overbar{\bd})} :\nnR^N & \rightarrow & \pR \\
		\pwrv       & \mapsto     & \left\{
    \begin{array}{ll}
    \underset{m \in \setM}{\text{min}} \dfrac{\overbar{R}_n p_n}{W\log_2\begin{pmatrix}1+s_n(\pwrv,\overbar{\bw},\overbar{\bd},m)\end{pmatrix}} & \text{if } p_n > 0 \\
    \underset{m \in \setM}{\text{min}} \dfrac{\overbar{R}_n \ln{2}}{W h_{m,n}} \left (  I_{m,n'} +  \sigma^2_{\text{noise}} \right ) & \text{otherwise,} \\
    \end{array}\right.
	\end{array}
$}
\end{equation*}
for every $n \in \setN$.
Note, that $t_n^{(\overbar{\bw},\overbar{\bd})}$ is a positive concave mapping with continuous extension at $p_n = 0$ that fulfills the properties of \thref{def_2} given in the Appendix.

Consequently, for $(\overbar{\bw},\overbar{\bd}) \in \setQ$ and a maximum power budget $\overbar{P} > 0$, the utility maximization problem in \eqref{eq11plus} can be stated as the power allocation problem:
\begin{subequations} \label{eq12}
	\begin{alignat}{3}
		  & \underset{\pwrv,c}{\text{\normalfont{maximize}}} & \quad & c \tag{\ref{eq12}}                                                                                                                  \\
		  & \text{\normalfont{subject to}}                   &       & \pwrv \in \text{Fix}\left ( cT^{(\overbar{\bw},\overbar{\bd})} \right )\label{eq12a}                                                    \\
		  &                                     &       & \left \| \pwrv \right \|_\infty  \le \overbar{P}                                                                          \label{eq12b} \\
		  &                                     &       & \pwrv \in \nnR^N, c \in \pR .   \label{eq12c}
	\end{alignat}
\end{subequations}

The problem in \eqref{eq12} is a particular case of \thref{problem_1} in the Appendix. It can be solved with the simple iterative fixed point algorithm given in \eqref{eq19} in the Appendix.
Its relation to problem \eqref{eq11} can summarized as follows.
Suppose that $(\bw^{\star},\bd^{\star})$ is the optimal beam configuration to problem \eqref{eq11}.
If we solve \eqref{eq12} by fixing $\overbar{\bw} = \bw^{\star}$ and $\overbar{\bd} = \bd^{\star}$, then the solution $(c^{\star},\pwrv^{\star})$ to \eqref{eq12} is also the optimal fraction $c^{\star}$ and power $\pwrv^{\star}$ to problem \eqref{eq11} .
Furthermore, the entries of the optimal \ac{UE}-\ac{AP} assignment vector $\boldsymbol{m}^{\star} = \left ( a_1^{\star},...,a_N^{\star} \right ) \in \mathcal{M} \times \cdots \times \mathcal{M}$ can be recovered via
\begin{equation}\label{ue_ap_assignment_equation}
m^{\star} \in \underset{m \in \setM}{{\argmin}}  \hspace{3pt} \dfrac{\overbar{R}_n p^{\star}_n}{W\log_2\begin{pmatrix}1+s_n(\pwrv^{\star},\bw^{\star},\bd^{\star},m)\end{pmatrix}}.
\end{equation}
As shown above, if the optimal beam configuration is known, \eqref{eq11} can be solved optimally with a simple algorithm. 
\subsection{Receive Beam Width and Direction Adjustment using Simulated Annealing Heuristics} \label{s6b}
Now, we propose a meta-heuristic based on \ac{SA} \cite{haggstrom2002finite} to obtain the optimal beam configuration.
Recall that the \ac{SA} algorithm works with a parameter called temperature $\tau$, which is to be cooled down as the beam configurations change.
The notion of cooling is interpreted as decreasing the probability of accepting solutions with worse \emph{utility} as the search space is explored.
We define the following main components of the \ac{SA} that are relevant to our optimization problem:
\begin{enumerate}[leftmargin=13pt]
    \item \textbf{Solution presentation:} The solution presentation for $(\bw,\bd)$ determines that the \emph{utility} $U(\bw,\bd)$ (obtained by solving \eqref{eq12} with $\overbar{\bw} = \bw$ and $\overbar{\bd} = \bd$) is associated with the beam width and direction adjustment problem.   
    \item \textbf{State transition mechanism (neighborhood search):} The algorithm starts from the initial state $(\bw_{\text{init}},\bd_{\text{init}}) \in \setQ$. The state $(\bw_{\text{init}},\bd_{\text{init}})$ is chosen such that all \acp{AP} select the largest beam width possible and the direction that points to the hotspot area. The main idea of the neighborhood search is that for a given temperature $\tau$, we randomly select a new state $\left( \bw',\bd' \right) \in \setQ \setminus \left( \bw,\bd \right)$, calculate the corresponding \emph{utility} \eqref{utility_map}, and replace the current solution $\left(\bw,\bd \right)$ with $\left(\bw',\bd' \right)$ if the \emph{utility} is improved.
    \item \textbf{Cooling procedure:} At the initial stage, the \ac{SA} algorithm starts with the highest possible temperature, $\tau_{\text{max}}$.
    Throughout an iterative procedure, the temperature is gradually decreased.
    In each iteration and for a given temperature $\tau$, the algorithm determines $\Delta U = U(\bw',\bd') -U(\bw,\bd)$ and computes the acceptance probability $\text{Pr}(\Delta U)$ of the new solution:\vspace{10pt}
    \begin{equation}\label{eq17}
        \text{Pr}(\Delta U) = \left\{\begin{array}{ll}
        e^{\frac{\Delta U}{\tau}} &, \Delta U \le 0\\
        1 &,  \Delta U > 0
        \end{array}\right.
    \end{equation}    
    In case of $\Delta U>0$ the new solution is always accepted.
    For $\Delta U \le 0$ the new solution accepted if $\text{Pr}(\Delta U) > rand(0,1)$.
    This scheme aims to jump out from a temporary local minimum.
    The acceptance probability of the new solution decreases as the temperature decreases or as the \emph{utility} of the new state is insufficient ($\Delta U$ obtains a large negative value) as shown in \eqref{eq17}. 
    \item \textbf{Termination criteria:} The \ac{SA} algorithm terminates if no improvement on the \emph{utility} is reached after a certain number of iterations. Otherwise, it continues the search procedure until the final temperature is reached.
\end{enumerate}
The implemented steps of our \ac{SA} scheme are given in Algorithm~\ref{alg1}.
After termination, we obtain a triplet $(\pwrv^\star,\bw^\star,\bd^\star)$ that solves the problem in \eqref{eq11}, and corresponding \ac{UE}-\ac{AP} assignment can be recovered from solution, as shown in \eqref{ue_ap_assignment_equation}.
\renewcommand{\algorithmicrequire}{\textbf{Input:}}
\renewcommand{\algorithmicensure}{\textbf{Output:}}
\setcounter{algorithm}{1}
\begin{algorithm}[!t]
	\caption{Receive Beam Width and Direction Adjustment}
	\label{alg1}
	\begin{algorithmic}[1]
		\REQUIRE $\overbar{P}, \setQ , \bw_{\text{init}},\bd_{\text{init}}, i_{\text{max}}, \tau_{\text{max}}, \tau_{\text{min}},$
		\ENSURE $(\pwrv^\star, \bw^\star, \bd^\star)$
		\renewcommand{\algorithmicensure}{\textbf{Initialize:}}
		\ENSURE $\tau = \tau_{\text{max}}, \bw = \bw_{\text{init}}, \bd = \bd_{\text{init}}$
		\STATE Compute \emph{utility} \eqref{utility_map} for $(\bw,\bd)$
		\WHILE{$\tau > \tau_{\text{min}}$}
		\FOR{$i=1$ \TO $i_{\text{max}}$}
		\STATE Compute \emph{utility} \eqref{utility_map} for $(\bw',\bd') \in \mathcal{Q} \setminus \left( \bw,\bd \right)$
		\STATE $\Delta U = U(\bw',\bd') -U(\bw,\bd)$
		\STATE flag = 1
		\IF{$\Delta U>0$}
		\STATE Accept $(\bw,\bd) \leftarrow (\bw',\bd')$
		\ELSE
		\STATE Calculate probability, $\text{Pr}(\Delta U) = e^{\frac{\Delta U}{\tau}}$
		\IF{$\text{Pr}(\Delta U) > rand(0,1)$}
		\STATE Accept $(\bw,\bd) \leftarrow (\bw',\bd')$
		\ELSE
		\STATE Reject $(\bw,\bd) \leftarrow (\bw,\bd)$
		\STATE flag = 0
		\ENDIF
		\ENDIF
		\IF{flag = 1}
		\STATE Update $(\bw^\star,\bd^\star) = (\bw,\bd)$
		\ENDIF
		\ENDFOR
		\vspace{3pt}
		\STATE $\tau = \tau / \log(i+1)$
		\vspace{3pt}
		\ENDWHILE
		\STATE $(\pwrv^\star, c^\star) \leftarrow$ solution to \eqref{eq11plus} with $(\bw^\star,\bd^\star)$
	\end{algorithmic}
\end{algorithm} 
\vspace{-15pt}
\section{Numerical Evaluation}\label{s7}
\subsection{Millimeter-wave Propagation Model}\label{s7a}
We use the \ac{mmWave} path loss model proposed in \cite{Rappaport2017}.
In this work, the directional antenna patterns and gains are adapted to the \ac{PL} model. 
The distance-dependent \ac{PL} function in [dB] is given as follows:
\begin{equation}\label{eq300}
\resizebox{0.89\hsize}{!}{$
	\begin{array}{ll}
		\text{PL}_{\lbrack\text{dB}\rbrack}(f_c,d)  =
		  & \text{FSPL}(f_c,d_0) + 10 \alpha \log_{10}(d) + X_{\sigma},
	\end{array}
	$}
\end{equation}
where $d$ is the transmission distance in meters, $\text{FSPL}(f_c,d_0)$ is the free space path loss for carrier frequency $f_c$ in GHz at reference distance $d_0$, $\alpha$ is the path loss exponent and $X_{\sigma}$ is a zero mean Gaussian random variable with standard deviation $\sigma_{\text{SF}}$ in dB (shadowing).
It is a common assumption to set $d_0 = 1$ m.
As described in \cite{Rappaport2017}, the above model can be parametrized for the so-called urban micro (UMi) open square \ac{LoS} scenario.
For an applicable range of $6 < f_c < 100$ GHz, we then obtain:
\vspace{-3pt}
\begin{equation}\label{eq301}
\resizebox{0.89\hsize}{!}{$
	\begin{array}{ll}
		\text{PL}_{\lbrack\text{dB}\rbrack}(f_c,d) = 32.4 + 18.5 \log_{10}(d) + 20 \log_{10}(f_c) + X_{\sigma}
	\end{array}
	$}
	\vspace{-3pt}
\end{equation}
This scenario refers to high user density open areas with \ac{AP} heights below rooftop (approx. $20$ m), \ac{UE} heights at ground level (approx. $1.5$ m) and a shadow fading of $\sigma_{\text{SF}}=4.2$ dB.
\begin{table}[b]
	\begin{center}
		\normalsize	
		\resizebox{0.95\hsize}{!}{$
		\begin{tabularx}{1\linewidth}{l l}
            \Xhline{2\arrayrulewidth}
			\textbf{Parameters}                              & \textbf{Value}    \\
			\hline
            \vspace{-10pt} \\
			\ac{UE} (Tx) number $N$                          & 20                \\
			\ac{AP} (Rx) number $M$                          & 3                 \\	
		    Inter-site shadowing correlation                 & 0.5               \\	
            Carrier frequency $f_c$                          & 28 GHz            \\
			System bandwidth $W$                             & 1 GHz             \\
			Noise power density    & -145 dBm/Hz       \\
			Sidelobe gain $\epsilon$                         & 0.1               \\
			\ac{UE} beam widths $\theta^{\text{Tx}}$         & $90^o$           \\
			\ac{UE} beam directions $[ \beta^{\text{Tx}}_{\text{min}},\beta^{\text{Tx}}_{\text{max}} ]$ & $[ 250^o,290^o ]$     \\
			\ac{AP} beam widths $\mathbb{D}_{\theta}$        & $\{30^o,45^o,60^o\}$                 \\
			\ac{AP} beam directions $\mathbb{D}_{\beta}$     & $\{70^o,80^o,90^o,100^o,110^o\}$     \\ [0.5ex]
			\Xhline{2\arrayrulewidth}
		\end{tabularx}
		$}
	\end{center}
	\caption{Basic simulation parameters.}	
	\label{tab1}
    \vspace{-15pt}
\end{table}
\begin{figure}[t]
	\hspace{-5pt}
	\includegraphics[width=85mm]{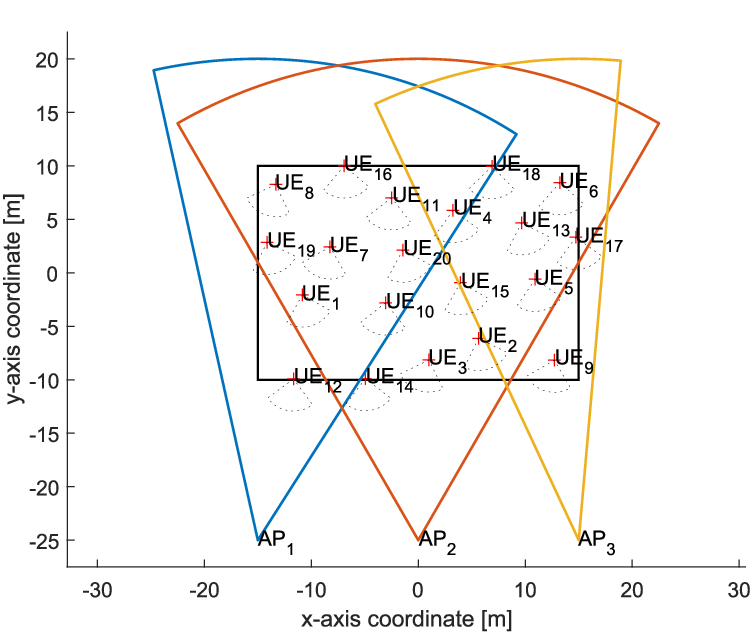}
	\caption{Network layout used for the simulations (shown for the beam configuration $\bw = (45^o,60^o,30^o)$, $\bd = (80^o,90^o,100^o))$.}
	\label{fig10}
    \vspace{-11pt}
\end{figure}
\subsection{Simulation Setup} \label{s7b}
\begin{figure}[t]
\vspace{0pt}
	\centering
	\includegraphics[width=96mm]{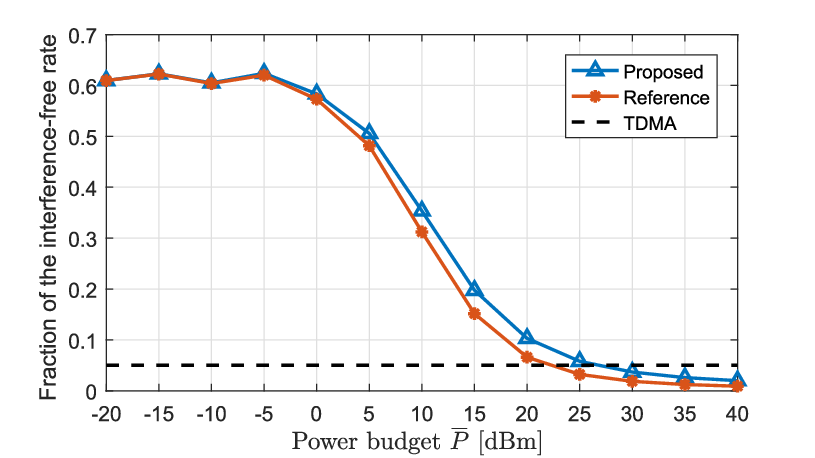}
	\caption{Allocated fraction of the interference-free rate over increasing power budget per user, corresponding to weight $c$ in the \emph{proposed} scheme and to the \ac{UE} with minimum fraction in the \emph{reference} scheme.}
	\label{fig7c-1}
	\includegraphics[width=96mm]{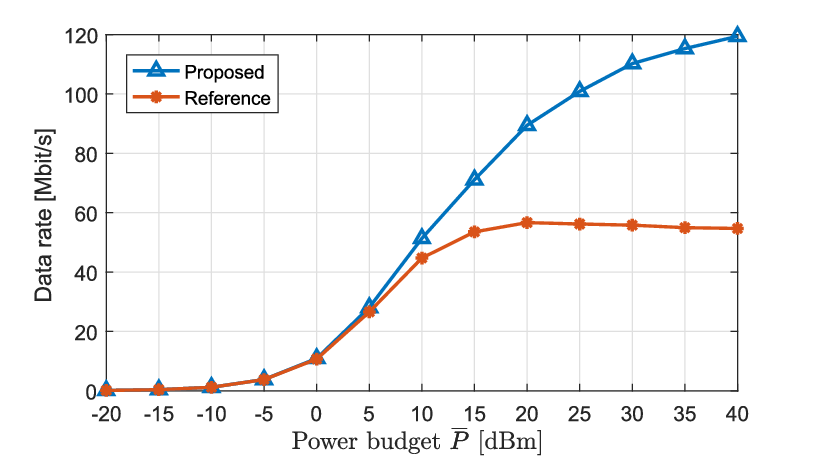}
	\caption{Uplink rate of the \ac{UE} with minimum allocated fraction over increasing power budget per user.}
	\label{fig7c-2}
	\includegraphics[width=96mm]{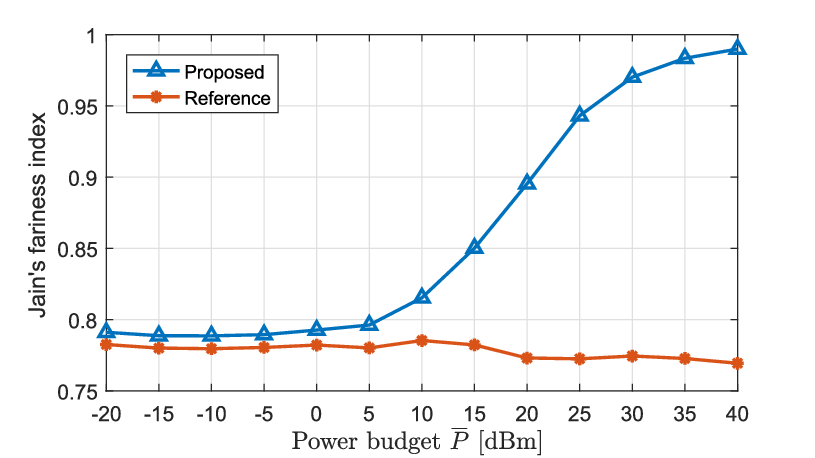}
	\caption{Evaluation of the fairness in the network through Jain's fairness index $\mathcal{J}=\frac{(\sum_{n=1}^N R_n)^2}{N \sum_{n=1}^N R_n^2}$ applied to the uplink rates over increasing power budget per user.}
	\label{fig7c-3}
	\vspace{-11pt}
\end{figure}
\begin{figure*}[ht]
	\centering
	\includegraphics[width=185mm,center]{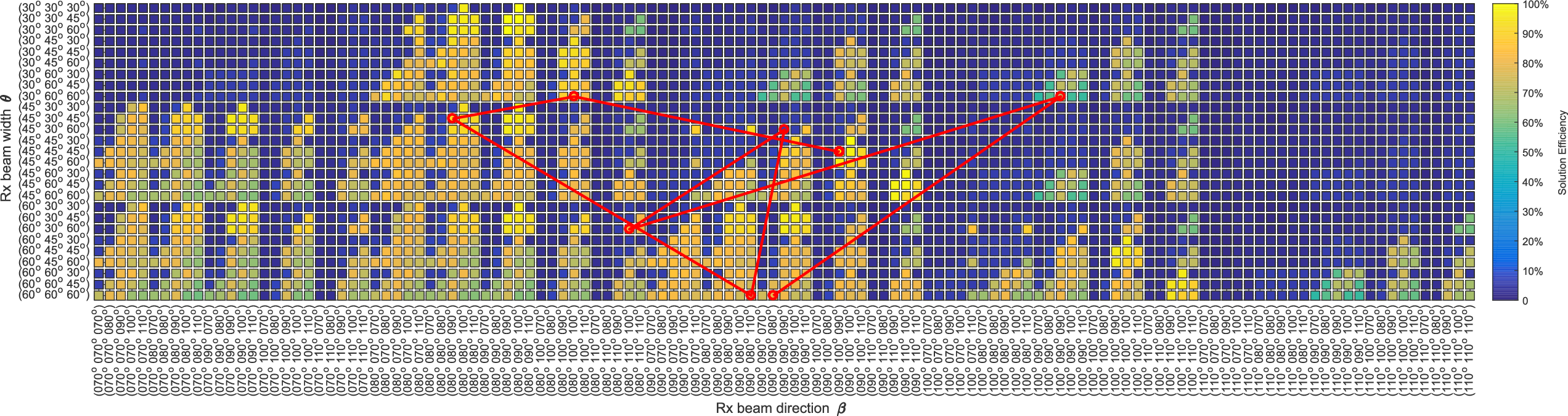}
    \caption{Simulated annealing (SA) performance for $\overbar{P} = 30$ dBm using half of the search space ($\tau_{\text{max}} = 42$, $i_{\text{max}}=42$) compared to brute force (BF). The efficiency is $98.3397 \%$ where $100 \%$ is the global optimum by the BF solution. The outcomes (arguments) of the BF and SA solutions are: $\bw^{\text{BF}} = (30^o,30^o,30^o)$ and $\bd^{\text{BF}} = (80^o,90^o,100^o)$, $\bw^{\text{SA}} = (45^o,45^o,45^o)$ and $\bd^{\text{SA}} = (90^o,100^o,90^o)$ }
	\label{fig7d-1}
    \vspace{-20pt}
\end{figure*}
\begin{figure}[ht]
	\centering
	\includegraphics[width=95mm]{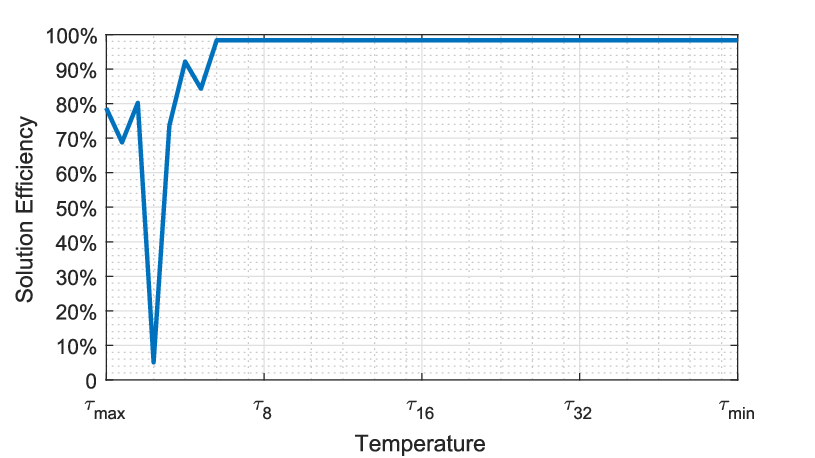}
	\caption{Solution efficiency of the SA algorithm as a function of decreasing temperature $\tau$.}
	\label{fig7d-2}
    \vspace{-11pt}
\end{figure}
\begin{figure}[ht]
	\hspace{-12pt}
	\includegraphics[width=96mm]{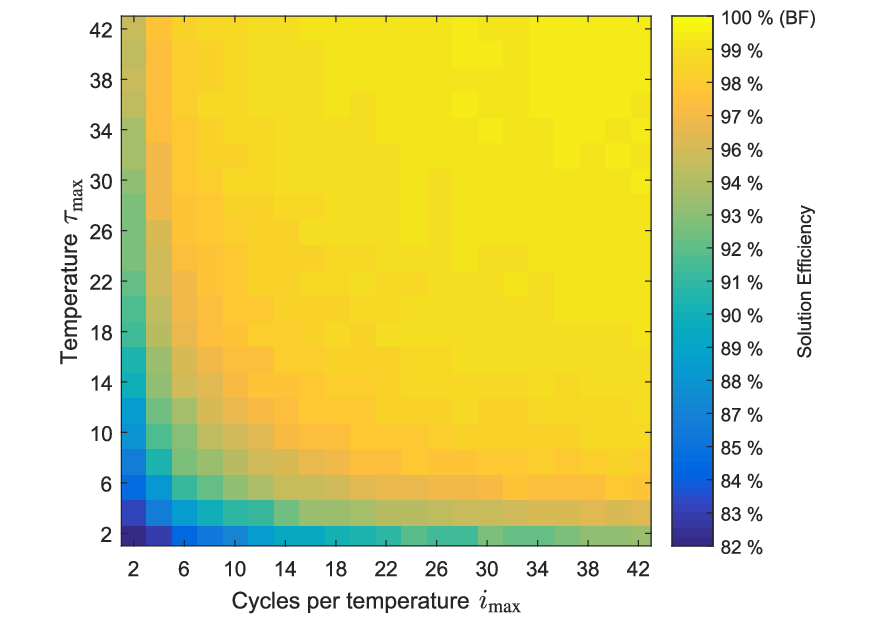}
    \caption{SA performance for $\overbar{P} = 30$ dBm with different parameters $\tau_{\text{max}}$ and $i_{\text{max}}$ in Algorithm 2. The parametrization influences the search space in the iterations ($\tau_{\text{max}} = 42$, $i_{\text{max}}=42$ corresponds to half search space as compared to BF)}
	\label{fig7d-3}
    \vspace{-10pt}
\end{figure}

For the performance evaluation of our proposed method, we consider a \ac{mmWave} access network with \acp{UE} that are distributed uniformly at random within a hotspot area considering a separation distance of $4$ m.
The size of the area is $30 \times 20$ m${}^2$ and the \ac{AP} locations are as shown in Fig.~\ref{fig10}.
Further system parameters are listed in Table~\ref{tab1}.
The simulations are averaged over 500 random realizations of user positions.

\subsection{Simulations} \label{s7c}
In Fig. \ref{fig7c-1} to Fig. \ref{fig7c-3}, we give the results from our \emph{proposed} scheme using fixed point algorithms for a specified beam configuration $\bw = (45^o,60^o,30^o)$, $\bd = (80^o,90^o,100^o)$.
For comparison, we also show the performance of a \emph{reference} scheme, which assumes fully interfered transmissions with maximum power $\overbar{P}$ of each \ac{UE}.
It can be seen that, in the noise-limited power regime, both schemes perform similarly while our proposed approach outperforms the full power transmission in the interference-limited range.
Not only is the worst network user (with minimum allocated fraction) made better off (Fig.~\ref{fig7c-2}), but the overall fairness in the network is also largely increased (Fig.~\ref{fig7c-3}).
In addition, Fig. \ref{fig7c-1} shows that in \ac{mmWave} networks, at a certain operation point in terms of $\overbar{P}$, schemes that utilize orthogonal resources such as time-division multiple access (TDMA) may be preferable than schemes treating interference as noise (TIN), for the reason that TDMA guarantees the constant rates for all \acp{UE}.
In TDMA, the fraction of the interference-free rate can be simply given as $c \le 1/N$.
As a general outcome from the study of our simulation setup it can be stated that interference cannot be ignored in our particular \ac{mmWave} scenario. 

Below, we show the improvements of the \emph{proposed} weighted rate allocation scheme after running the \acf{SA} heuristic in Algorithm 2.
First, we exemplify the performance weighted rate allocation scheme over the whole beam configurations set $Q$ in Fig. \ref{fig7d-1}, i.e., for all possible receive beam configurations with the discrete candidate sets $\mathbb{D}_{\theta}$, $\mathbb{D}_{\beta}$.
We use the small size problems since we compare the \ac{BF} solution (in a large search space \ac{BF} solution becomes an infeasible) to know the global optimum. 
In particular, we show the relative performance, called solution efficiency, compared to the best solution of problem \eqref{eq11} when a brute force (BF) search is applied (denoted by $100 \%$ solution efficiency).
The red path in Fig. \ref{fig7d-1} marks the neighborhood search with jumps in the states (beam configurations) when the utility is improved.
Fig. \ref{fig7d-2} shows how the utility develops in the cooling procedure as parameter $\tau$ decreases over several iterations.
Finally, we illustrate the solution efficiency in the simulated scenario when parameters in Algorithm 2 are changing. 
There is a trade-off between temperature $\tau$ and number of cycles $i_{\text{max}}$ per temperature which impact the utility.
Hence, a certain parametrization can be obtained for a desired operational point.  

%
%

\section{Conclusion}\label{s8}
In this work we have proposed an interference management and beamforming mechanism for uplink hotspot \ac{mmWave} communication on shared resources.
In particular, our centralized scheme jointly optimizes the uplink power, the \ac{UE}-\ac{AP} assignments, and the receive beam configurations of the \acp{AP}. The proposed approach combines a simple fixed point algorithm with a heuristic based on \ac{SA}, which is used to search for the optimal beam configurations. We showed that, if the \ac{SA} heuristic is able to find the optimal beam configuration, then the fixed point algorithm provides us with the optimal power and the \ac{UE}-\ac{AP} assignments. Nevertheless, even if the beam configuration produced by the \ac{SA} heuristic is a suboptimal beam configuration, then the fixed point algorithm is still optimal in the sense of maximizing the common fraction of interference-free rates for the given beam configuration.   



\appendix
The results in this study are related to properties of \acp{SIF}, which are defined as follows:
\begin{definition}\thlabel{def_2}
A function $f:\nnR^N \rightarrow \pR$ is said to be a standard interference function if the following properties hold:
\begin{enumerate}
\item (Scalability) $(\forall \boldsymbol{x} \in \nnR^N)$ $(\forall \alpha > 1)$ $\alpha f(\boldsymbol{x}) > f(\alpha \boldsymbol{x})$
\item (Monotonicity) $(\forall \boldsymbol{x}_1 \in \nnR^N)$ $(\forall \boldsymbol{x}_2 \in \nnR^N)$ $\boldsymbol{x}_1 \ge \boldsymbol{x}_2 \Rightarrow $ $f(\boldsymbol{x}_1) \ge f(\boldsymbol{x}_2)$
\end{enumerate}
If $T : \nnR^N \rightarrow \pR^N$ is given by $(\forall \boldsymbol{x} \in \nnR^N) \hspace{5pt} T(\boldsymbol{x})=[f_1(\boldsymbol{x}),...,f_N(\boldsymbol{x})]$, where $f_i:\nnR^N \rightarrow \pR$ are \acp{SIF} for every $i \in \{ 1,...,N \} $, then $T$ is said to be a standard interference mapping.
\end{definition}
It is known that positive concave functions (e.g., the coordinate-wise functions used to construct $T$ in \eqref{eq12}) are a subclass of \acp{SIF} \cite{Cavalcante2016, Renato2018globalSIP}.
Furthermore, if $T : \nnR^N \rightarrow \pR^N$ is a standard interference mapping, the following optimization problem, which is a generalization of that in \eqref{eq11}, can be solved optimally with a simple normalized fixed point algorithm that we show below \cite{Renato20162, Nuzman2007}. 
\begin{problem}[Canonical form of the utility maximization problem]\thlabel{problem_1}
\begin{subequations} \label{eq18}
	\begin{alignat}{3}
		&   \underset{\boldsymbol{x},c}{\text{{\normalfont maximize}}} & \quad & c \tag{\ref{eq18}}                                                                                                   \\
		&   \text{{\normalfont subject to}}  &  &  \boldsymbol{x} \in \text{{\normalfont Fix}}(cT):=\left \{ \boldsymbol{x} \in \nnR^N \mid \boldsymbol{x} = cT(\boldsymbol{x}) \right \}  \label{eq18a}   \\
		  &   &   & \left \| \boldsymbol{x} \right \|_l  \le \overbar{X} & \label{eq18b} \\
		  &   &   & \boldsymbol{x} \in \nnR^N,  c \in \pR            & \label{eq18c}
	\end{alignat}
\end{subequations}
\end{problem}
\noindent where  $\overbar{X} \in \pR$ is a design parameter (e.g., a power budget $\overbar{P}$), $\left \| \cdot \right \|_l$ is an arbitrary monotone norm, and $T:\nnR^N \rightarrow \pR^N$ is an arbitrary continuous concave mapping in the class of standard interference mapping.
In particular, the vector $\boldsymbol{x}^{\star}$ that solves \thref{problem_1} is the limit of the sequence $\left ( \boldsymbol{x}_n \right )_{n\in\mathbb{N}}$ generated by \cite{Nuzman2007}
\begin{equation}\label{eq19}
\begin{array}{lr}
    \boldsymbol{x}_{n+1} = T'(\boldsymbol{x}_n):= \dfrac{\overbar{X}}{\left \| T(\boldsymbol{x}_n) \right \|_l }T(\boldsymbol{x}_n),& \boldsymbol{x}_0 \in \nnR^N.
\end{array}
\end{equation}
Once $\boldsymbol{x}^{\star} = \lim_{n \rightarrow \infty } \boldsymbol{x}_n$ is known, we recover the optimal scalar $c^{\star}$ of \thref{problem_1} by $c^{\star}:=\dfrac{\overbar{X}}{\left \| T(\boldsymbol{x}^{\star}) \right \|_l }$.


\section*{Acknowledgment}

This work is partially supported by grant no. 01MD18008B of the German Federal Ministry for Economic Affairs and Energy (DigitalTWIN project).

\begin{acronym}[SPACEEEEEE]
	\acro{SIF}{standard interference function}
	\acro{3GPP}{3rd generation partnership project}
	\acro{AP}{access point}
	\acro{AWGN}{additive white Gaussian noise}
	\acro{BF}{beamforming}
	\acro{CSI}{channel state information}
	\acro{CSI}{channel state information}
	\acro{D2D}{device-to-device}
	\acro{FSPL}{free space path loss}
	\acro{GHz}{gigahertz}
	\acro{INR}{interference-to-noise ratio}
	\acro{ISI}{intersymbol interference}
	\acro{IoT}{Internet of Things}
	\acro{LoS}{line-of-sight}
	\acro{M2M}{machine-to-machine}
	\acro{MAC}{medium access control}
	\acro{MIMO}{multiple-input multiple-output}
	\acro{MTC}{machine-type communication}
	\acro{MT}{mobile terminal}
	\acro{MUSA}{multi-user shared access}
	\acro{NOMA}{non-orthogonal multiple access}
	\acro{NORA}{non-orthogonal resource allocation}
	\acro{OFDM}{orthogonal frequency division multiplexing}
	\acro{PHY}{physical layer}
	\acro{PL}{path loss}
	\acro{QoS}{quality-of-service}
	\acro{QuaDRiGa}{Quasi Deterministic Radio Channel Generator}
	\acro{RB}{resource block}
	\acro{RMS}{root mean square}
	\acro{RRM}{radio resource management}
	\acro{SER}{symbol-error rate}
	\acro{SINR}{signal-to-interference-plus-noise ratio}
	\acro{SIR}{signal-to-interference ratio}
	\acro{SNR}{signal-to-noise ratio}
	\acro{UE}{user equipment}
	\acro{cMTC}{critical machine-type communications}
	\acro{dB}{decibel}
	\acro{eMBB}{enhanced mobile broad-band}
	\acro{mmWave}{millimeter-wave}
	\acro{nLoS}{non-line-of-sight}
	\acro{SA}{simulated annealing}
	\acro{CF}{cost function}
\end{acronym}

\bibliographystyle{IEEEtran}
\bibliography{references} 
\end{document}